\documentclass[sigconf,nonacm]{acmart}

\AtBeginDocument{%
  }

\copyrightyear{2026}
\acmYear{2026}
\setcopyright{cc}
\setcctype{by}
\acmConference[KDD '26]{Proceedings of the 32nd ACM SIGKDD Conference on Knowledge Discovery and Data Mining V.1}{August 09--13, 2026}{Jeju Island, Republic of Korea}
\acmBooktitle{Proceedings of the 32nd ACM SIGKDD Conference on Knowledge Discovery and Data Mining V.1 (KDD '26), August 09--13, 2026, Jeju Island, Republic of Korea}
\acmDOI{10.1145/3770854.3783932}
\acmISBN{979-8-4007-2258-5/2026/08}

\usepackage{caption} 
\usepackage{subcaption}
\usepackage{graphicx}
\newcommand{\indep}{\perp \!\!\! \perp}

\begin{document}

\title{Choice Modeling and Pricing for Scheduled Services}


\author{Adam N. Elmachtoub}
\affiliation{%
  \institution{Columbia University and Amazon}
  \city{New York City}
  \state{NY}
  \country{USA}
}
\email{machtoub@amazon.com}

\author{Kumar Goutam}
\affiliation{%
  \institution{Amazon}
  \city{Seattle}
  \state{WA}
  \country{USA}
}
\email{kggoutam@amazon.com}

\author{Roger Lederman}
\affiliation{%
  \institution{Amazon}
  \city{Seattle}
  \state{WA}
  \country{USA}
}
\email{rllederm@amazon.com}

\renewcommand{\shortauthors}{Adam N. Elmachtoub, Kumar Goutam, and Roger Lederman}

\begin{abstract}

We describe a novel framework for discrete choice modeling and price optimization for settings where scheduled service options (often hierarchical) are offered to customers, which is applicable across many businesses including some within Amazon. In such business settings, the customers would see multiple options, often substitutable, with their features and their prices. These options typically vary in the start and/or end time of the service requested, such as the date of service or a service time window. The costs and demand can vary widely across these different options, resulting in the need for different prices. We propose a system which allows for segmenting the marketplace (as defined by the particular business) using decision trees, while using parametric discrete choice models within each market segment to accurately estimate conversion behavior. Using parametric discrete choice models allows us to capture important behavioral aspects like reference price effects which naturally occur in scheduled service applications. In addition, we provide natural and fast heuristics to do price optimization. For one such Amazon business where we conducted a live A/B experiment, this new framework outperformed the existing pricing system in every key metric, increasing our target performance metric by 19\%, while providing a robust platform to support future new services of the business. The model framework has now been in full production for this business since Q4 2023.

\end{abstract}


\keywords{decision trees; choice models; pricing; service options offerings; business analytics}

\maketitle
\pagestyle{empty}
\section{Background}

There are many businesses that require customers to choose among options for a scheduled service. For instance, a customer may pick a day (also called lead time) and/or time window for delivering their groceries, for a technician to come to their location, or to bring a vehicle to a repair center. Accordingly, different days or time windows (and lengths) may have different prices in some of these applications. In this paper, we describe how to formulate such a problem, leverage customer choice data properly, and price the different service options. Our framework has been deployed successfully within Amazon, and this paper provides an overview of the approach and the results from our implementation. 

\subsection{Customer Journey}
We begin by describing the problem from a customer's perspective. The standard customer experience for any of such businesses would look like the following: many customers would go to the business website, enter a few details of their requirements, and then expect to see prices of various options which would fulfill their needs and requirements. These prices need to be shown fairly quickly after entering the details, so latency is a crucial aspect. The Amazon business we worked with shows the customer a set of options and their prices simultaneously. Notably, for some customers, some of the options might be considered substitutable. If the customer likes the price of a particular option, they select and book it (a \textit{conversion}). In many cases, there will also be a second level where more options are provided after selecting an option from the first level. The  second level options will provide more flexibility and details to match the user preferences, and these secondary options need to be priced as well. For this paper, we suppose that the first level option corresponds to a particular day (also called lead time), and the second level option is a time window within that day. 

An example of such a use-case is when customers are shopping on a grocery store website and they have to choose their desired delivery date and time. At the first level, the customer may see possible delivery dates a few days into the future (which are usually called lead-times). Once they click on a particular date, the customer will see possible delivery windows for that day. Each day and time may have a different delivery cost associated with it, which may need to be optimized based on cost and demand information.

Depending upon which time window the customer chooses, it will affect the cost of service for the business while providing value to the customer. Overall, our problem statement is to select the prices of all these options with a goal to maximize (mixture of) performance criteria. The goal of this work is to design a system that \textit{(i)} predicts conversion probabilities of the various options accurately, \textit{(ii)} optimizes the prices of those options to maximize a given business objective, and \textit{(iii)} handles new business capabilities such as pricing second (and third) level options.

\subsection{Legacy approach}

For the particular business where we implemented our proposed methodology, the previous legacy system for automating pricing was using the following ideas: first discretize prices, then fit a decision tree (on a bootstrapped dataset to navigate the exploration-exploitation trade-off  \citep{elmachtoub2017practical}) corresponding to each price to predict the conversion probability for that price, and then select the objective-maximizing price for each of the service options. The previous framework had three fundamental limitations which we describe below, motivating us to invent the next generation pricing system.

\begin{figure}[ht]
     \centering
     \begin{subfigure}[b]{0.48\textwidth}
         \centering
         \includegraphics[width=\textwidth]{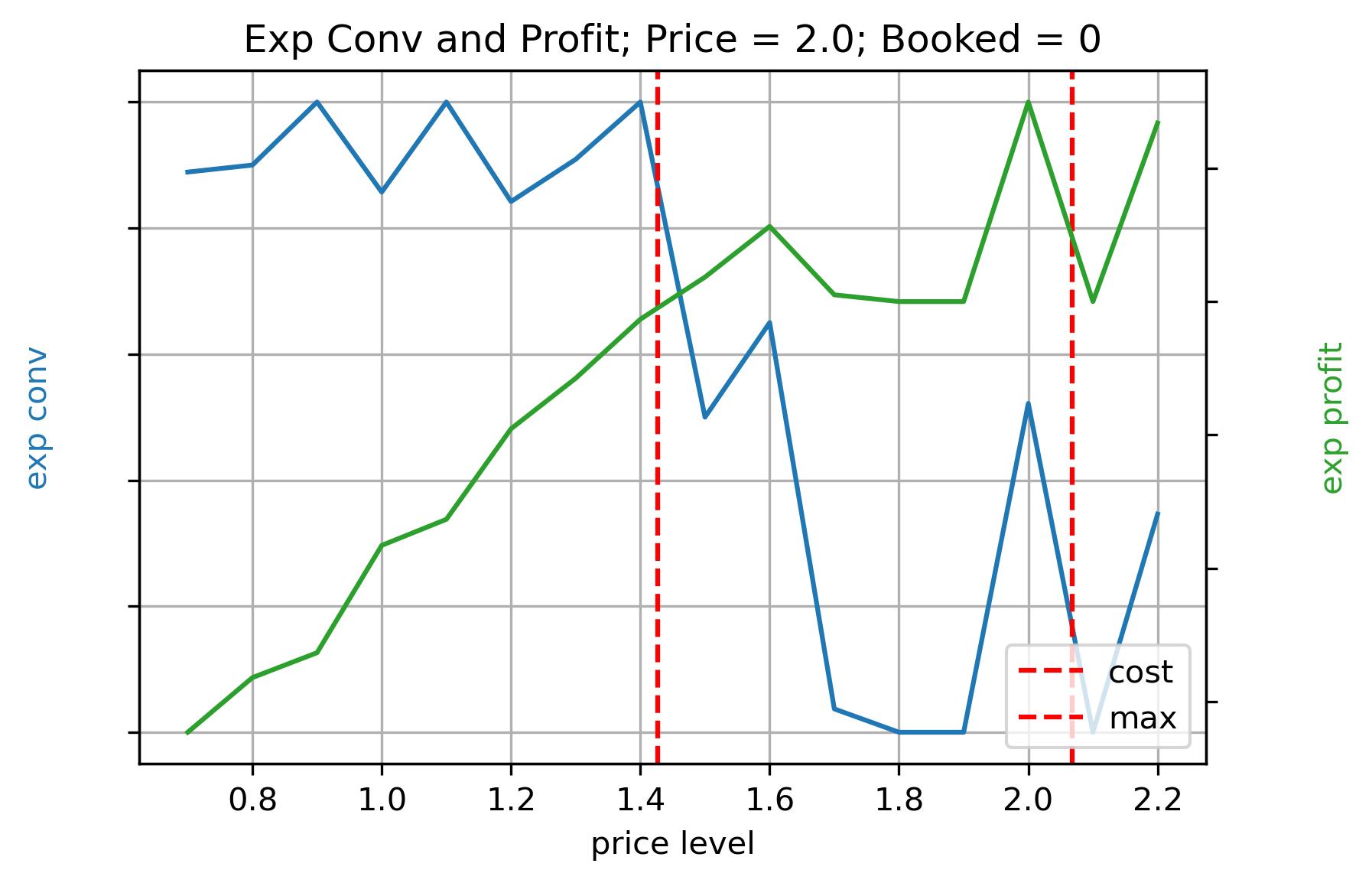}
         \caption{ }
         \label{fig:quote}
     \end{subfigure}
     \hfill
     \begin{subfigure}[b]{0.48\textwidth}
         \centering
         \includegraphics[width=\textwidth]{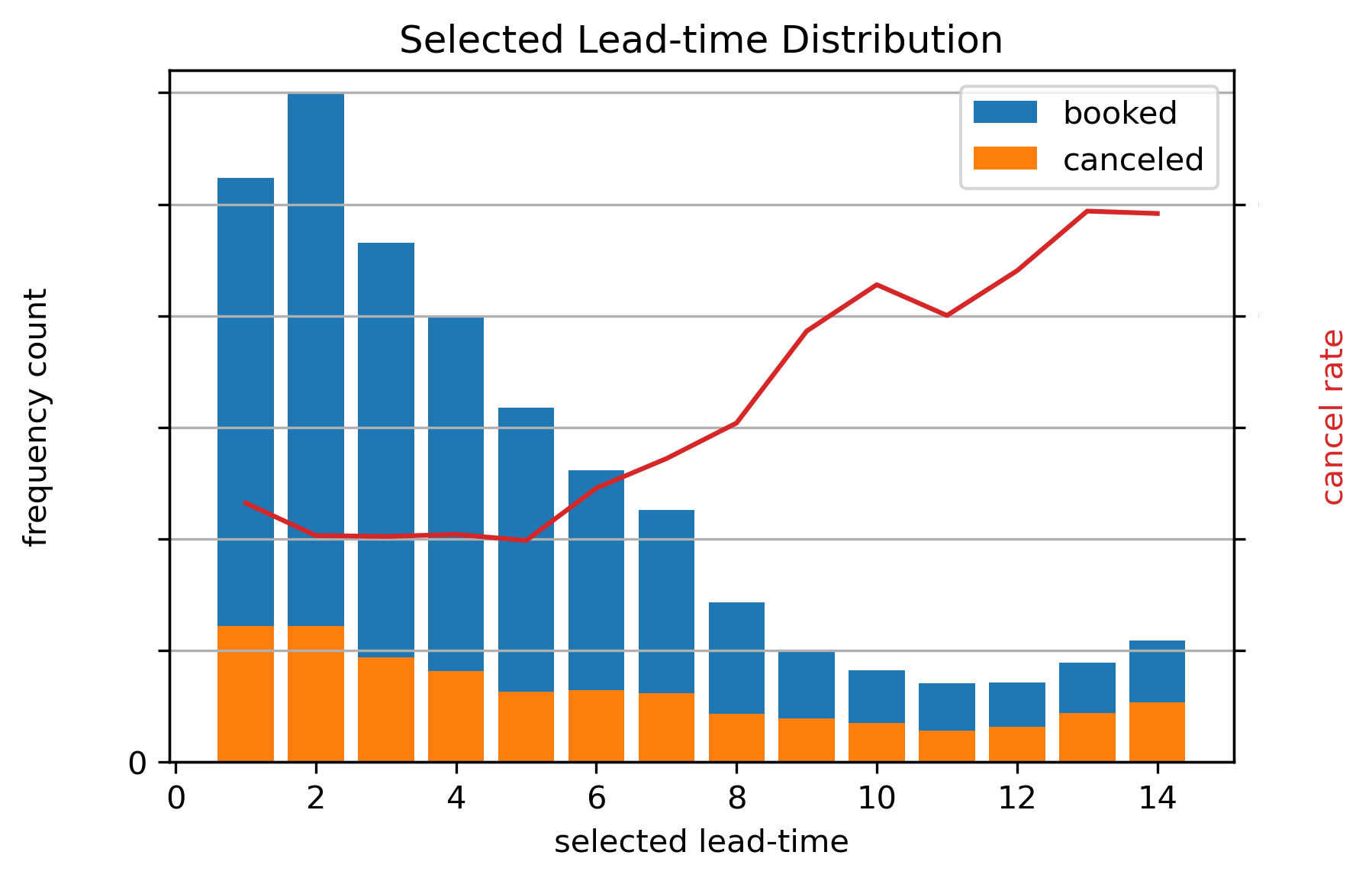}
         \caption{}
         \label{fig:motivation}
     \end{subfigure} 
     \caption{(a) A single price request example. We show the conversion probabilities and expected objective (averaged over 14 lead times) computed by the old model. The final price was 2.0 (scaled) and the customer did not book. The dotted red lines are price guardrails. (b) Price requests from a certain time range. We show cancellation rate, executed quotes, and canceled quotes as a function of selected lead time option for booked customers.}
     \Description{A single price request example}
\end{figure}

We first observe that the previous system does not leverage the data optimally in estimating the conversion probabilities since each price has its own data and machine learning model, rather than using all data for one unified model. As a result, the expected conversion rate as a function of price can have an erratic and non-monotone shape due to limited data at some price levels, which can in turn lead to selecting sub-optimal prices. See Figure \ref{fig:quote} for an anecdotal example. Note that the y-axes in figures in this paper are intentionally hidden and prices are normalized for confidentiality.

Second, we note that there is no substitutability captured across various options in the legacy framework, resulting in an algorithm that optimizes the price for each day (lead time) option separately. In other words, the legacy system does not explicitly capture the fact that customers may be inclined to switch their choice from one day to another, if the price is right. Figure \ref{fig:motivation} demonstrates that customers clearly have preferences for particular days (also called lead times), so a reasonable model would predict that customers may substitute from their most preferred day to the day before or the day after (as opposed to many days before or after the original lead time) for the right price. Even if the legacy framework was somehow extended to capture substitutability, optimizing multiple discrete prices  would have been an intractable combinatorial optimization problem.

Third, the previous framework did not easily allow us to consider new products or initiatives for future needs. For example, a business introduced differentiated pricing for new time window options, e.g,  different prices for specific pickup windows and/or delivery windows of different lengths. This second level of choice would be welcome by customers to have more control on pickup/delivery logistics, but increases learning and computational difficulty. Thus, a more robust system for learning and optimization was needed that can handle modifications and enrichment of the business.

\subsection{Our New (and Implemented) Methodology} 

We implemented a new model framework for choice modeling and price optimization that addresses the fundamental issues discussed above. 
This system can efficiently learn and tractably optimize prices for a potentially large set of substitutable service options, through a novel integration of both parametric and non-parametric models. The key components of this framework are as follows.

\begin{itemize}
\item \textbf{Market Segmentation Tree.} First, the marketplace itself is segmented into different segments, based on the type of pricing quote request, using the Market Segmentation Tree (MST) methodology \citep{aouad2023market}. An MST is a decision tree of the marketplace based on differences in choice behavior rather than differences in the pricing quote characteristic, i.e., it is a supervised approach to market segmentation based on customer response data. This leverages the flexibility of non-parametric tree methods, but allows for structural models of choice behavior to be integrated at each leaf node of the decision tree.

\item \textbf{Reference-Price-Effects Choice Model.} In each market segment, a modified multinomial logit (MNL) discrete choice model with local reference price effects is used to model how customers in this market segment decide between the various (lead time) options according to the prices offered. The use of the parametric MNL choice model naturally allows us to capture substitutability among the lead time options in each market segment. The reference price effect here is essential and captures customer's willingness to adjust their desired lead time $\pm1 $ day  for the right price. As one can see from the pattern in Figure \ref{fig:motivation}, customers have a clear range of days (lead times) in mind when making a decision. 

\item \textbf{Assortment Pricing Algorithms.} We provide optimization procedures to quickly find prices of the entire assortment, i.e., all lead time (and other available) options, simultaneously. Our algorithms leverage the structure of the reference-price-effects MNL discrete choice model \cite{wang2018prospect} to come up with strong candidate solutions quickly. Specifically, we consider a simple policy that has a minimum price parameter and a cost-markup price parameter. 

\item \textbf{Flexible Wait Time Capability.} We build an auxiliary choice model to capture which time window (the second level) options the customers prefer, as a function of their prices and the lead time (the first level option) selected. We use natural assumptions to justify why we do not need to capture all combinations of time windows and lead times as individual choices, which ensures a fast, simple, as well as adaptable methodology.

\end{itemize}

\subsection{Literature Review} 
The idea of modeling and estimating customers' choice  behavior and using it in a revenue management setting has been explored in many industries (see \cite{garrow2007much,besbes2020pricing,fisher2014demand,ye2018customized,gallego2019revenue} for more details). Multiple classes of models can be used to estimate choice behavior (also known as the demand function): ranging from parametric models to non-parametric models. The most famous and widely used choice model is the Multinomial Logit (MNL) model \cite{train2009discrete}. A key feature of this choice model is that it has closed-form choice probabilities, can be fit using a convex Maximum Likelihood Estimation, and the optimal pricing strategy is a single constant mark-up across all options \cite{hopp2005product,wang2012capacitated}.

Non-parametric models such as those discussed in \cite{farias2013nonparametric, chen2022decision, aouad2022representing, wang2024neural, cai2022deep} can have their own advantages in terms of estimation accuracy and allowing for more automatic flexibility. In this paper, we use Market Segmentation Trees (MSTs) from  \cite{aouad2023market}, which involves an interpretable non-parametric tree that has choice models in the leaves. In our setting, we focus on using MNL choice models in the leaves with local reference price effects, inspired by \cite{wang2018prospect}. We elaborate on these ideas and how we extend them to our application in the remainder of the paper.

The rest of the paper is structured as follows. In Section \ref{sec:formulation}, we formally describe the problem and assumptions. In Section \ref{sec:choice}, we describe the use of the market segmentation trees along with the reference-price-effects choice model to predict customer behavior. In Section \ref{sec:optimization}, we describe the optimization algorithms.  In Section \ref{sec:implementation}, we describe the details of the A/B experiment for one particular Amazon business which generated about \textbf{19\% improvement in performance} for the new framework over the previous system. In Section \ref{sec:ext}, we describe how this new approach can be extended to new service offerings such as flexible time-windows. We provide conclusions in Section \ref{sec:conc}.

\section{Assortment Pricing Problem Formulation}\label{sec:formulation}

We now describe a general model framework for assortment pricing, which can be applicable to many businesses including several within Amazon. We let $\mathbf{X} \in \mathbb{R}^d$ denote the features corresponding to a customer and their pricing quote request, which could include business specific details and general temporal aspects such as day of week. Features may also include the current conditions of the Amazon network (relevant info regarding costs) and the economy (holiday season).

We let $\mathbf{C} \in \mathbb{R}^L$ denote the random cost associated with a pricing quote for each of the $L$ lead time options we are quoting. Let $\mathbf{p}\in \mathbb{R}^L$ denote the corresponding vector of price quotes, which is the decision variable. All price and cost data can be normalized by some benchmark or base price.
We let $Y(\mathbf{p})\in \{0,\ldots,L\}$ denote the lead time option selected by the customer as a function of $\mathbf{p}$, where $Y(\mathbf{p})=0$ indicates that the customer did not select any of the $L$ lead time options. Let $Z\in \{0,1\} $ be the event that the customer cancels their request after booking, i.e., $Z=1$ indicates a cancellation. Although we cannot reveal the exact business objective (which is a convex combination of expected revenue, expected cost and expected conversion), we have given a specific example of what the objective function would look like if one were to myopically maximize expected profit.
The profit objective would be, given a pricing request described by $\mathbf{x}$, to choose the prices so as to maximize the expected profit, i.e., 
\begin{align}
\max_{\mathbf{p} \geq 0} \mathbb{E}\left[\sum_{i=1}^L (p_i - C_i)\mathbb{I}(Y(\mathbf{p})=i)(1-Z) |\mathbf{X}=\mathbf{x}\right]  \label{eq:obj1}
\end{align}
Assuming that $\mathbf{C} \indep Y(\mathbf{p}) | X$ as well as $\mathbf{C} \indep Z  | X$ (we use $\indep$ to denote independence), then the objective expression above can be re-formulated using the following steps:
\begin{equation*}
\begin{aligned}
& \mathbb{E}\left[\sum_{i=1}^L (p_i - C_i)\mathbb{I}(Y(\mathbf{p})=i)(1-Z) |\mathbf{X}=\mathbf{x}\right] & \\
= & \sum_{i=1}^L \mathbb{P}(Y(\mathbf{p})=i|\mathbf{X}=\mathbf{x}) \cdot \\ 
& \mathbb{E}\Big[\sum_{j=1}^L (p_j - C_j) \mathbb{I}(Y(\mathbf{p})=j)(1-Z)  |\mathbf{X}=\mathbf{x}, Y(\mathbf{p})=i \big] \\
= & \sum_{i=1}^L \mathbb{P}(Y(\mathbf{p})=i|\mathbf{X}=\mathbf{x}) \cdot \\ 
& \mathbb{E}\Big[  (p_i - C_i) (1-Z) |\mathbf{X}=\mathbf{x}, Y(\mathbf{p})=i \Big] \\ 
= & \sum_{i=1}^L \mathbb{P}(Y(\mathbf{p})=i|\mathbf{X}=\mathbf{x}) \cdot \\ 
& \Big[ (p_i - \mathbb{E}\big[ C_i|\mathbf{X}=\mathbf{x}, Y(\mathbf{p})=i)  \big]) \mathbb{P}(Z=0|\mathbf{X}=\mathbf{x}, Y(\mathbf{p})=i)  \Big] \\
= & \sum_{i=1}^L \mathbb{P}(Y(\mathbf{p})=i|\mathbf{X}=\mathbf{x}) \cdot \\
& \Big[ (p_i - \mathbb{E}\big[ C_i|\mathbf{X}=\mathbf{x} \big]) \mathbb{P}(Z=0|\mathbf{X}=\mathbf{x}, Y(\mathbf{p})=i)  \Big]
\end{aligned}
\end{equation*} 
The first equation follows from the law of total expectation. The second equation follows by definition of the indicator function. The third equation follows from the assumption that $\mathbf{C} \indep Z  | X$. The fourth equation follows from the assumption that  $\mathbf{C} \indep Y(\mathbf{p}) | X$.

Thus, the assortment pricing problem can be reformulated as 
\begin{equation}
\begin{aligned} 
\max_{\mathbf{p} \geq 0} \sum_{i=1}^L \big[\mathbb{P}(Y(\mathbf{p})=i|\mathbf{X}=\mathbf{x}) & (p_i - \mathbb{E}\left[ C_i|\mathbf{X}=\mathbf{x} \right]) \cdot  \\ 
& \mathbb{P}(Z=0|\mathbf{X}=\mathbf{x}, Y(\mathbf{p})=i) \big] \label{eq:obj2}
\end{aligned}
\end{equation}

The expression \eqref{eq:obj2} above clearly shows that there is a need to come up with prediction models for  $C_i|\mathbf{X}=\mathbf{x}$,  $Z|\mathbf{X}=\mathbf{x},Y(\mathbf{p})=i$, and $Y(\mathbf{p})|\mathbf{X}=\mathbf{x}$. We note that we assume we have access to models to predict $\mathbb{E}\left[ C_i|\mathbf{X}=\mathbf{x} \right]$, which is not the focus of this paper. Predicting the cancellation rate, $\mathbb{P}(Z=1|\mathbf{X}=\mathbf{x},  Y(\mathbf{p})=i)$ is a straightforward classification task. The key challenge is designing a discrete choice model for $Y(\mathbf{p})|\mathbf{X}=\mathbf{x}$, which we detail in Section \ref{sec:choice}. 

We remark that removing the cost component in  \eqref{eq:obj2} results in a revenue maximization problem, while removing $(p_i - \mathbb{E}\left[ C_i|\mathbf{X}=\mathbf{x} \right])$ results in a conversion maximization problem. Our techniques apply to any of these objectives, as well as any convex combination of them. For completeness, we describe here the generalized objective as well (dropping the choice and cancellation events for simplicity):
\begin{align*}
\max_{\mathbf{p} \geq 0} \quad (1-\alpha)*\mathbb{E}\left[\sum_{i=1}^L (p_i - C_i)\right] + \alpha*\mathbb{E}\left[\sum_{i=1}^L C_i\right] 
\end{align*}
We can see that $\alpha=1$ maximizes expected weighted conversion, $\alpha=0.5$ maximizes expected revenue, and $\alpha=0$ maximizes expected profits. At the same time, one can choose any value $0\leq \alpha \leq 1$ depending upon the business requirements to balance revenue, cost and conversion priorities which gives flexibility to achieve the desired business goals.  

\section{New Solution Framework} \label{sec:choice}

In this section, we explain our new system which combines the Market Segmentation Tree (MST) and a local reference-price-effects multinomial logit (MNL) choice model to predict customer behavior. Recall that our goal is to to estimate the probability that a customer chooses lead time option $i; 0\leq i \leq L$ for any quote $\mathbf{x}$ and prices $\mathbf{p}$, i.e., $\mathbb{P}(Y(\mathbf{p})=i|\mathbf{X})$.

We shall estimate this probability using two steps. First, we build a decision tree by partitioning on the features in $\mathbf{X}$. We use the Market Segmentation Tree (MST) methodology in \cite{aouad2023market}, which explicitly builds trees that lead to the best choice models in the leaves. Within each leaf node $l$, we train a multinomial logit choice model with local reference price effects to capture $\mathbb{P}_l(Y(\mathbf{p})=i)$.

\subsection{Market Segmentation Tree}\label{sec:MST}
An MST is a type of binary decision tree where the tree is used to segment the marketplace, and choice models are fit in each leaf of the tree. The key idea behind an MST is that the splitting criteria is determined by the fit of the choice models in each leaf, resulting in a supervised way to do marketplace segmentation. We will think of each leaf as a market segment. The features $\mathbf{X}$ will be used to decide which leaf/market segment of the tree  we are in. Each leaf is associated with a particular choice model whose data and parameters are segment-specific. A flexible open-source library from \cite{aouad2023market} is available for MST in Python. The key advantage of using an MST is that we can capture nonlinear effects of the features using a non-parametric tree model, while still using a parametric model to do pricing and choice modeling within each segment. See Figure \ref{fig:mst} for an example. We describe the discrete choice models fit in each leaf in Section \ref{sec:ref} below.

\begin{figure}[ht]
  \centering
 \includegraphics[width=3.5in]{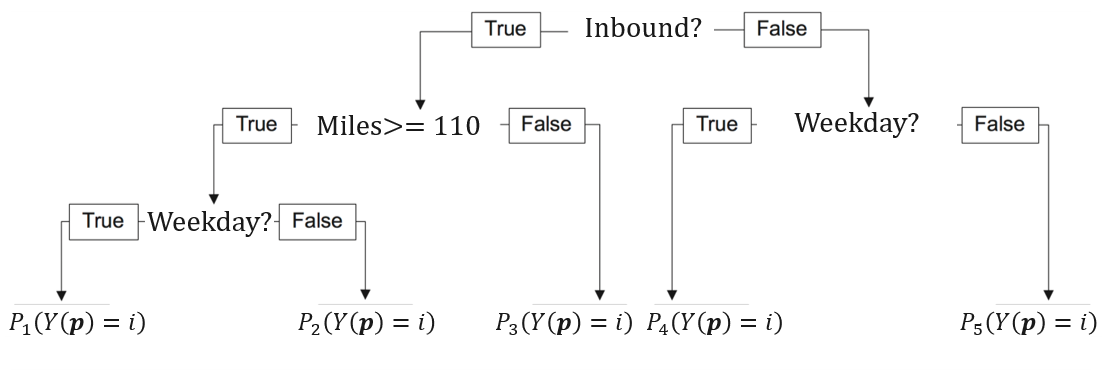}
     \caption{An example of an MST. Decision tree splits are performed with respect to the contextual variables provided, which includes both discrete and continuous features. 
     Each of the resulting market segments contains a unique MNL discrete choice model $\mathbb{P}_l(Y(\mathbf{p})=i)$.}
    \label{fig:mst}
    \Description{An example of an MST}
 \end{figure}

We remark that, in production, we can the train MST offline and update periodically. We have also deployed a step to subsample the data over the last few months to encourage exploration. This sub-sampling helps ensure we do not converge to a bad decision tree by introducing some variability and adapts to changing data over time. The intuition behind sub-sampling the dataset is akin to a technique known as Thompson Sampling in the multi-armed bandit literature \citep{elmachtoub2017practical}. Combining exploration and exploitation is fundamental as the ground truth fundamentally changes over time as the marketplace conditions change; thus, some exploration is needed to ensure adaptation occurs.

\subsection{Reference-Price-Effects Choice Model} \label{sec:ref}
\begin{figure}[ht]
     \centering
     \begin{subfigure}[b]{0.48\textwidth}
         \centering
         \includegraphics[width=\textwidth]{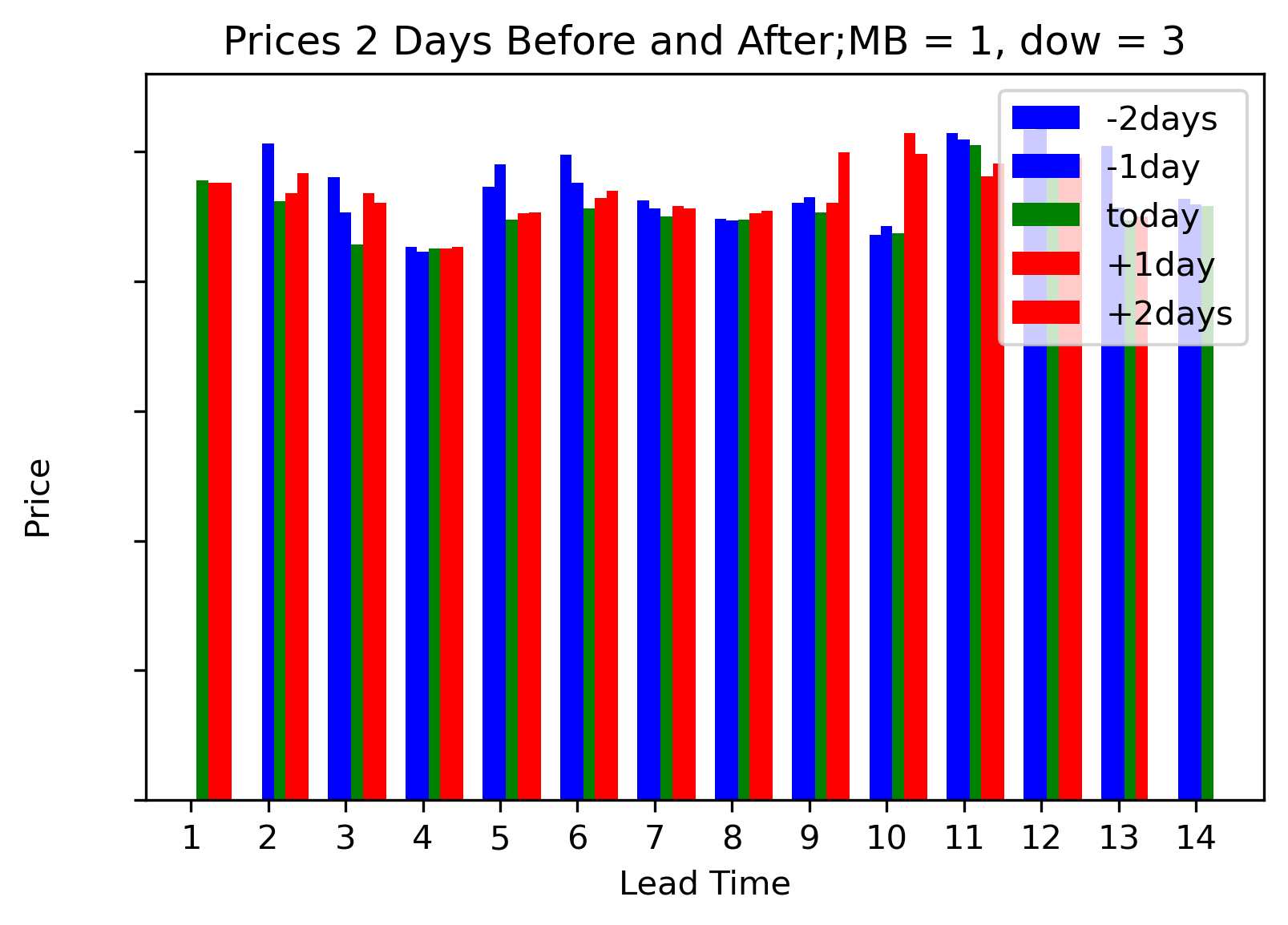}
         \caption{}
         \label{fig:2day}
     \end{subfigure}
     \hfill
     \begin{subfigure}[b]{0.48\textwidth}
         \centering
         \includegraphics[width=\textwidth]{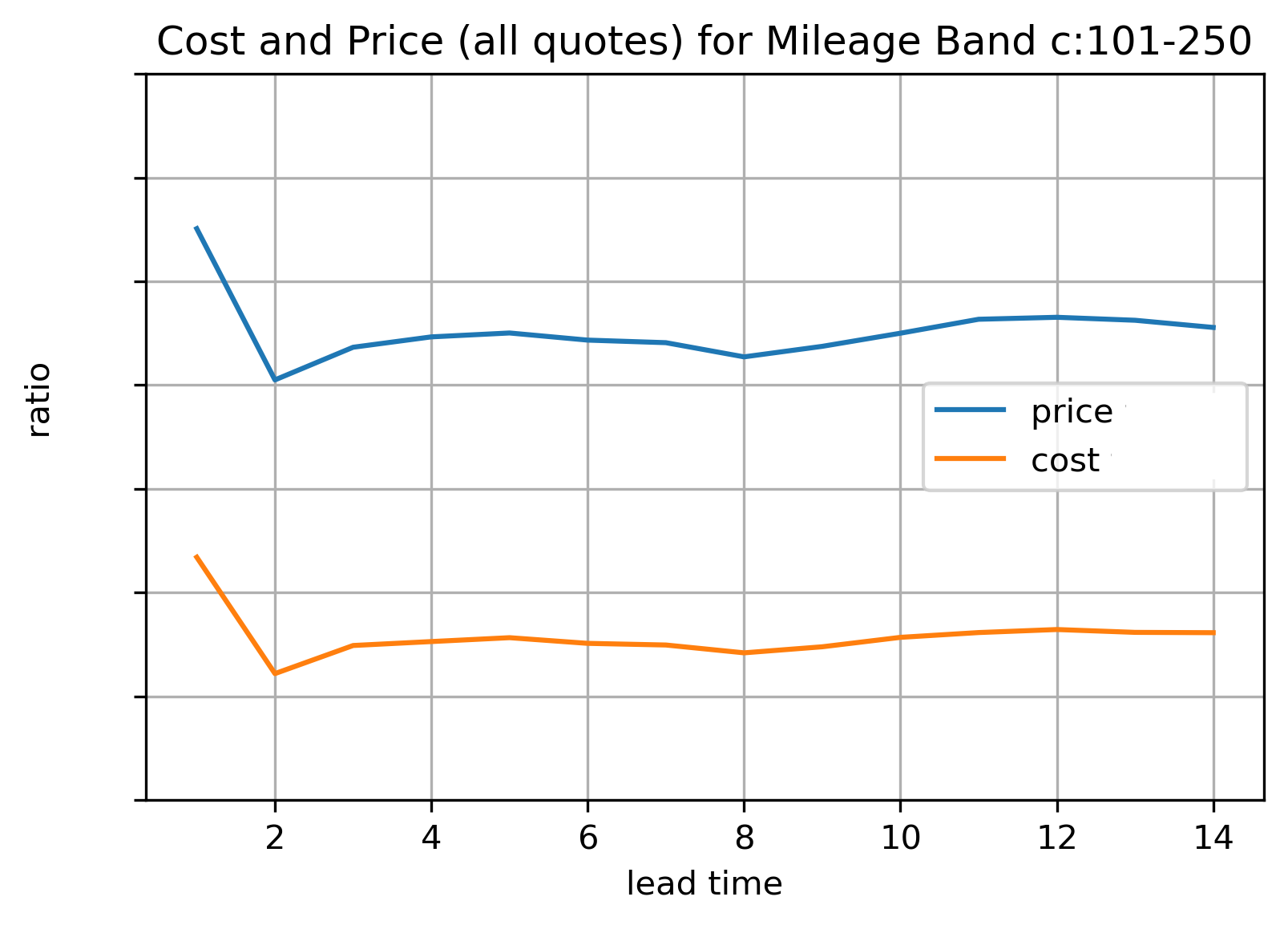}
         \caption{}
         \label{fig:cost}
     \end{subfigure} 
     \caption{(a) Filtered on booked quotes that were booked on Wednesday, we plot the average prices booked across all lead times. We also show the average prices $\pm 2$ days. (b) The average cost and price for the quotes as a function of lead time in the previous system.}
     \Description{Details}
\end{figure}

For the rest of the section, we focus on a particular leaf/market segment of our MST and drop the dependence on the leaf $l$ for convenience. The local reference price $r_i$ for lead time $i$ is defined as the lowest prices over a 3 weekday window centered at $i$ if $i$ is a weekday (2 days if $i$ is first or last weekday), or the lowest price over the weekend days if $i$ is on the weekend. From Figure \ref{fig:motivation}, we can see customers clearly have preferences towards particular lead times. However, Figure \ref{fig:2day} makes it clear that customers do have some leeway in their desired lead time and tend to favor the lower price among their search range. This motivates our usage of a local reference price effect, which helps capture substitution effects primarily to lead times that are close to the most preferred lead time.

We model the probability of a customer selecting lead time $i$ as 
\begin{align*}
    \mathbb{P}(Y(\mathbf{p})=i)= \frac{e^{\alpha_1 i + \alpha_2 i^2 + \alpha_3 \sqrt{i}  - \beta_i p_i - \gamma_i (p_i - r_i)}} {1+ \sum_{j=1}^L e^{\alpha_1 j + \alpha_2 j^2 + \alpha_3 \sqrt{j} - \beta_j p_j - \gamma_j (p_j - r_j)}},
\end{align*}
where $\alpha_1, \alpha_2, \alpha_3$ are constants to capture the lead time effects as seen in Figure \ref{fig:motivation}.  $\beta_i$ is the price sensitivity for lead time $i$, and $\gamma_i$ is the reference price sensitivity of lead time $i$ for having to pay more than the reference price. One can easily see that if we decrease price $p_i$, this would decrease the probability we choose any other option besides $i$. Moreover, if lead time $i$ has a price lower than $i+1$ (both weekdays or both weekends), then increasing the price of $i$ also provides an additional boost to the chance of choosing $i+1$ due to the reference price effect. This is ideal as the lead times closest to $i$ should be impacted the most by a change in the price of lead time $i$.

We now illustrate the mechanics of the reference-price-effects choice model with a toy example. Suppose there are 7 lead time options so that $L=7$ and the prices on those days are set to $p_1=\$10, p_2=\$11, p_3=\$9, p_4=\$12, p_5=\$12, p_6=\$10, p_7=\$8$ representing Sunday to Saturday prices. Following the recipe for local reference prices stated early, the reference prices are $r_1=\$8, r_2=\$9, r_3=\$9, r_4=\$9, r_5=\$10, r_6=\$10, r_7=\$8$. Suppose $\alpha_1=\alpha_2=\alpha_3=0, \beta_i=0.10$, and $\gamma_i=0.05$ for all $i=1,\ldots 7$. Then, $\mathbb{P}(Y(\mathbf{p})=i)$ is 0.295, 0.098, 0.089, 0.120, 0.076, 0.080, 0.109, 0.133 for $i=0,\ldots 7$ using the MNL formula above. Note that these probabilities sum to 1, and that option 5 has a higher probability than option 4 despite having the same price, since option 5 has a higher reference price. Option 3 has a higher probability than option 4 despite having the same reference price, since option 3 has a lower price. 

To estimate the model parameters $\alpha_1, \alpha_2, \alpha_3, \beta_i, \gamma_i$, we can use a standard technique know as maximum likelihood estimation (MLE), for which built-in packages are easily available for MNL based discrete choice models. Note that we have the historical prices, reference prices, and choices made to properly use MLE and fit our model. We also note that we can vary the definition of the reference prices: we can allow some dependencies on $\mathbf{X}$, and we can allow coefficients to have particular structure corresponding to weekend/weekdays and bucketing different lead times together. In Section \ref{sec:implementation}, we bucket the lead times into 4 categories to reduce the number of parameters so as to prevent over-fitting. In one of the use-cases where the business offers a 2 week period into the future, we split the 14 lead time options as follows: the first three weekdays, the next three weekdays, the last four weekdays, and the remaining four weekend days. 
 
We remark that one can have a very rich contextual MNL to describe the whole marketplace and not use MST, but this would require estimating too many parameters. The MST approach is an effective way to combine non-parametric interpretable methods like decision trees with parametric ones like MNL. The advantage of using the MNL model with local price reference effects is that it allows for tractable estimation, captures the behavioral effects of customers having a range of lead times in mind, and allows for price optimization, as described in Section \ref{sec:optimization} below.

\section{Price Optimization} \label{sec:optimization}

We first provide some background that is useful for our core price optimization problem \eqref{eq:obj2}. The paper \cite{wang2018prospect} considers the MNL model with a particular reference price effect. They show that if (i) the reference price is global, i.e., $r_i=\min_{j} p_j$, and (ii) all the price sensitivities $\beta_i$ are equal, then the optimal policy has a very special and tractable form: it has only two parameters corresponding to a minimum price and an optimal markup. Under the policy, each lead time option is priced at the maximum of (i) the minimum price and (ii) its cost plus the optimal markup.

Given that we do not have a global reference price, we have unequal price sensitivities, and we have other complex factors in our optimization model, the two-parameter policy is not necessarily optimal for our model. Nevertheless, it is still a reasonable and intuitive heuristic which we shall employ. 

Figure \ref{fig:cost} indeed suggests that some form of a markup policy is natural and aligns with current system. Our price for lead time option $i$ under the two parameter policy has the form 
\begin{align}
     p_i = \max \{m_1,  \quad c_i+1/( \beta_i+ \gamma_i)+m_2  \}, \label{eq:markup}
\end{align}
where $m_1$ is the minimum price, $m_2$ is the markup, $c_i$ is the expected cost, $\beta_i$ is the price coefficient for lead time $i$, and $\gamma_i$ is the reference price coefficient. 

The two-parameter policy has the following nice properties, convenient for this Amazon business' pricing application. First, it is very quick to optimize over due to its simplicity, which helps in keeping the latency low. Second, it keeps the prices of the lead time options in correspondence with their costs, which is essential for encouraging customers to use low cost options. Third, the small differences in cost estimation result in reference price effects, where customers end up pivoting to slightly lower prices but at a higher conversion rate. Finally, the minimum price ensures that if one of the costs is very low compared to the other products, we do not have a huge price gap. 

We subject this new framework to the same pricing guardrails as previously employed. These guardrails ensure that we earn a minimum margin, but do not unnecessarily charge customers a price too different from the going marketplace rate. Next, in Section \ref{sec:implementation}, we provide our results using real data and A/B testing. 

\section{Real World Results from A/B Test} \label{sec:implementation}

\subsection{Implementation Details and Challenges}
In this section, we describe some details and challenges we faced related to the actual implementation of this new framework. Firstly, there were several limitations in the way the pricing requests' data was getting captured. Since modeling the reference prices require accurate knowledge of the calendar dates of the options, it was imperative that these dates were captured and stored accurately. We found time zone inconsistencies which initially led to ill-fitted models and hence sub-optimal, unexpected prices. We had to build robust data pipelines and infrastructure to capture the raw data accurately and cleanly so that it can be used as directly as possible without a lot of extra steps to process it. 

Although we did find an open-source MST package from the original MST paper, it only had a simple model for the leaf nodes. Modifying the open-source MST package for our use case of having discrete choice models in the leaf nodes had its own issues. The new framework needed to be general enough to capture fitting choice models based on business data where the set of options presented to the customers need not be always the same; e.g., depending upon availabilities, not all lead-times will always be available. We also ran into multiple ill-defined matrices while fitting the model due to missing and repeated data.

Another challenge was hyper-parameter tuning, particularly related to tree depth (3-5) and minimum sample size (100-400) in the leaf-nodes. We also had to figure out the historical window size (6-12 weeks) to be used for model training. All these needed to be tested rigorously multiple times using historical data and we built a simulator to do so. As hinted previously in Section \ref{sec:MST}, we added a sub-sampling step on the historical data before fitting the model and estimating its parameters. This served two purposes: since day-over-day, not a lot of data would change (dropping the last day and adding the new day), the estimated models would be very similar. Sub-sampling insures that the fitted models are somewhat different day-over-day and hence we will not get stuck with a randomly bad model. Secondly, given that we still want to allow for price exploration, if we had the same model, it will result in the same optimal prices for the same pricing requests.

\subsection{Experimental Results}
We describe the A/B experiment setup here for one of the businesses at Amazon where we have successfully deployed the model. We ran the test for 6 weeks in 2023. All automated quote pricing traffic was randomized at a quote request level without any interference, so that 25\% of them went to our new pricing model framework, while the remaining 75\% continued to be priced by the legacy model framework. The new model was trained and updated at a daily cadence,using the most recent few weeks of data and then sub-sampling a dataset of about 50\% size to estimate the model parameters.

We now briefly describe a few segments from the MST from a particular day. The first segment corresponds to the distance feature provided and falls in the high range; the second and third segments are within medium and short range, and within this, are further split by a geography feature.
We found that  price sensitivities tend to be lower for earlier lead time options, meaning that a small price change affects earlier lead time options less than later lead times. We also found that the local reference price effects are statistically significant. 

\begin{table}[ht!]
  \centering
  \begin{tabular}{lrrrr}
    \toprule
    Model & NLL & Delta \% &  Brier Score & Delta \% \\
    \midrule
    Naive Baseline & 24945 & - & 0.4429 & -  \\
        Legacy Framework & 24248 & -2.8\% & 0.4245 & -4.2\% \\
        Vanilla MNL & 23971 & -3.9\% & 0.4303 & -2.8\% \\
        New Framework &  23461 & \textbf{-5.9\% }& 0.4193 &  \textbf{-5.3\%} \\
    \bottomrule
    \\
  \end{tabular}
  \caption{Model Fit Accuracy Metrics. NLL stands for Neg Log Likelihood. The naive baseline simply uses the average historical conversion rate for each lead time. Old framework is the legacy model. Vanilla MNL fits an MNL but does not do any market segmentation or capture reference price effects.}
  \label{tab:pred_perf_metrics}
\end{table} 

We compare and show the predictive performance improvement of the new model against the old one on held-out test data using classic metrics such as Brier score (akin to MSE) and negative log likelihood (NLL) in Table \ref{tab:pred_perf_metrics}. We also compare against a naive baseline that ignores features entirely. We see substantial gains of our new framework over the legacy framework, specifically over 3\% improvement in NLL and 1\% improvement in Brier Score. However, the legacy framework also has issues with non-monotonicity in the price-demand relationship, instability, and no substitution effect properties. Thus, it is important to also assess business metrics which we discuss next.

\subsection{Business Metrics Comparison}
We now describe the performance comparison results of this new model framework against the legacy one for major business metrics. As mentioned previously, we subject this new model framework to the same pricing guardrails as previously employed for an aples-to-apples comparison. For the specific objective we chose to optimize, we saw a dramatic 19\% improvement, which was also statistically significant. A detailed analysis of these results from the A/B test was conducted to gauge the statistical significance of the changes in key business metrics. We used standard two sided t-tests and all the improvements in revenue, profit and conversion prediction came out to be highly statistically significant (with all p-values less than or equal to 0.01). We decided to dial up to 100\% shortly after the conclusion of the A/B test.

Apart from the main business metrics, we also saw other benefits, including significantly lowering price variations compared to the legacy system (30\% reduction), reduction in the latency of price quoting (40\% faster), 5\% improvement in prediction accuracy, and 5\% reduction in average booked costs. The methodology and results were deemed a wide success, improving business metrics and many aspects of the customer experience.

\section{Extension to Second Level Time Window Options}  \label{sec:ext}
 
One of the businesses has started offering a new service giving customers even more flexibility and choices. In the most common case, second level options are the particular time-window options of the selected lead-time day.
 
Naturally, this tends to incur different costs and thus should result in different prices. However, customers benefit as different time-windows can be more convenient.

\begin{figure}[ht!]
     \centering
     \begin{subfigure}[b]{0.49\textwidth}
         \centering
         \includegraphics[width=0.64\textwidth]{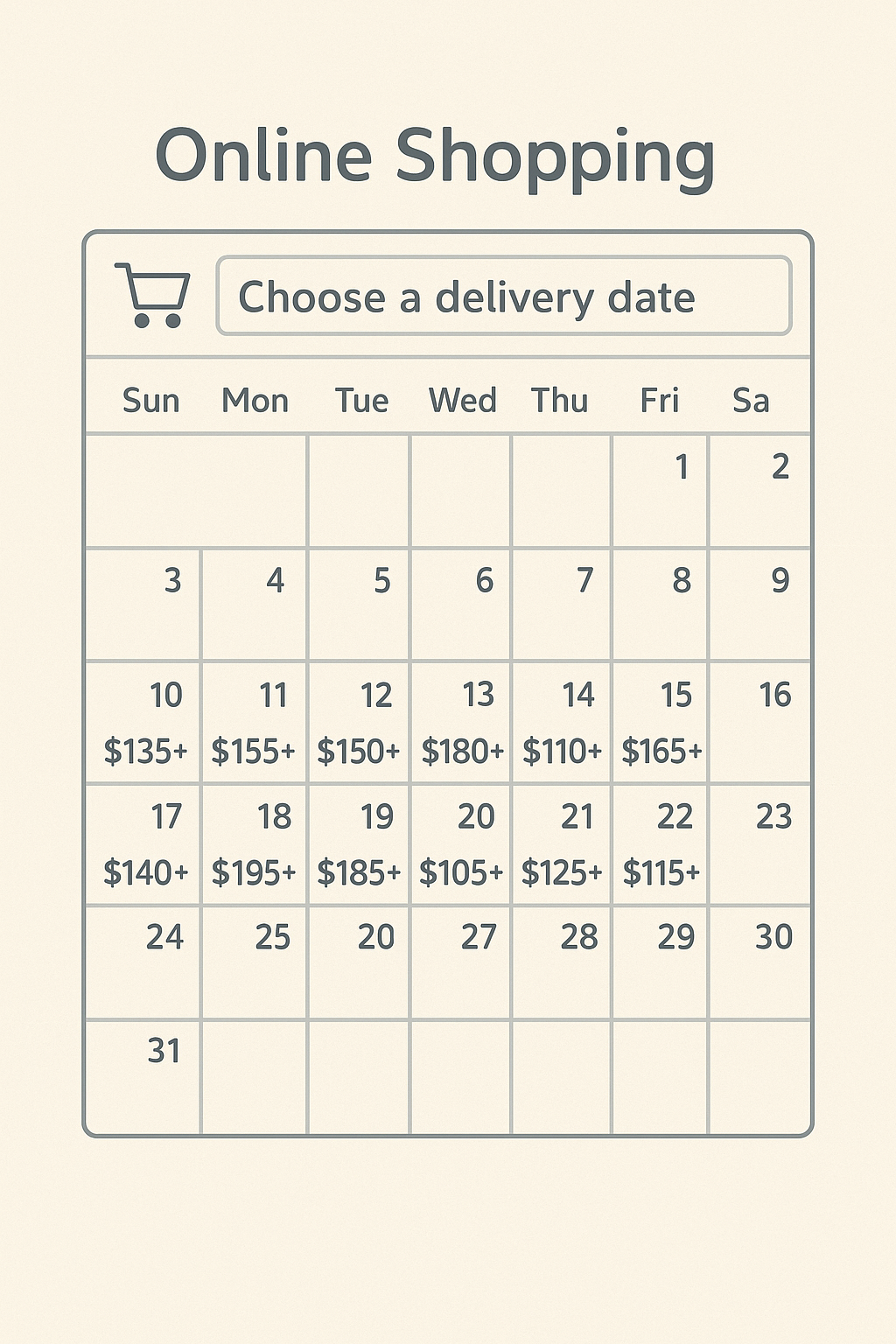}
         \caption{}
         \label{fig:leadtime}
     \end{subfigure}
     \hfill
     \begin{subfigure}[b]{0.49\textwidth}
         \centering
         \includegraphics[width=0.64\textwidth]{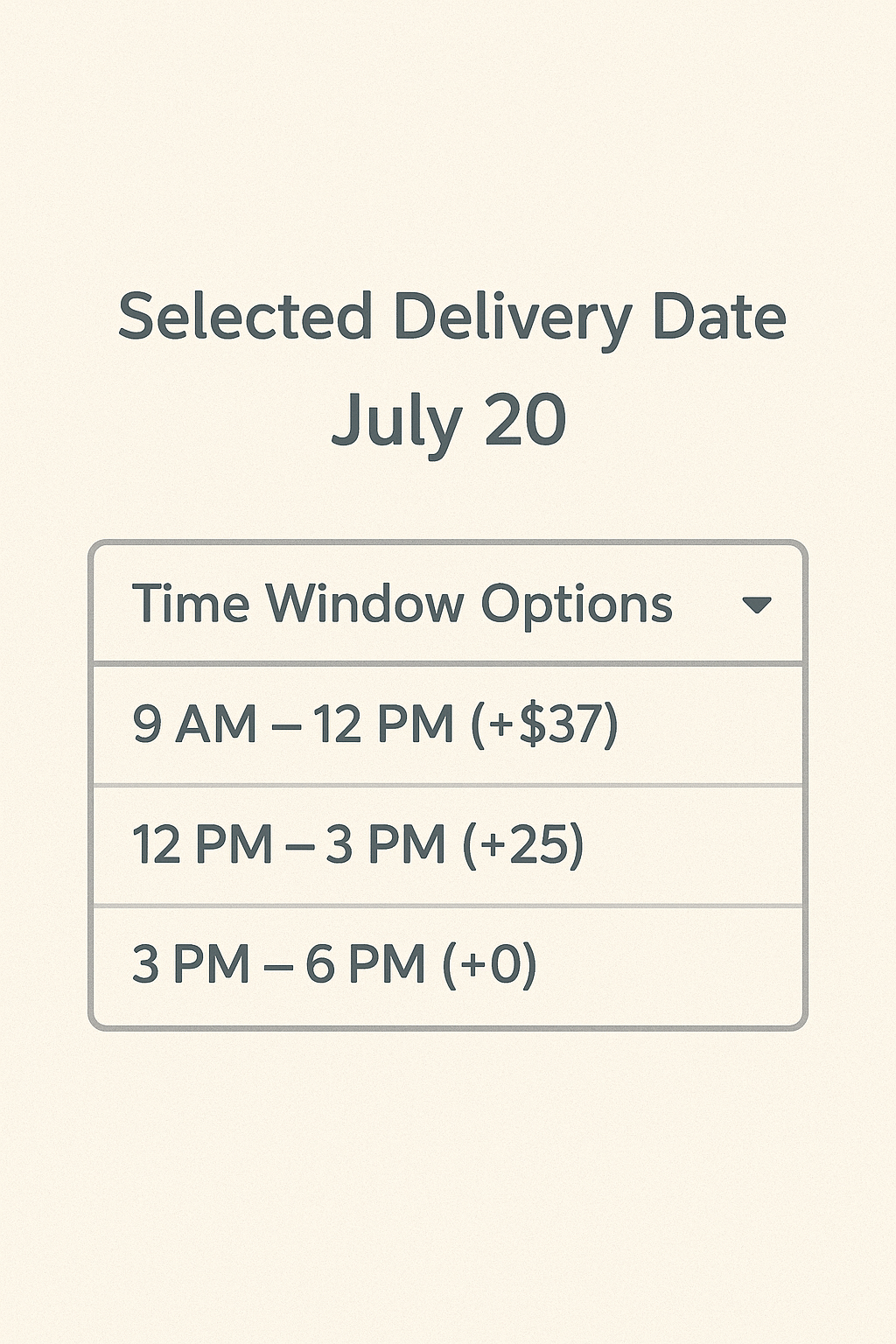}
         \caption{}
         \label{fig:shipwindow}
     \end{subfigure} 
     \caption{Lead Time and Time Window Options}
     \Description{Lead Time and Time Window Options}
\end{figure}

We provide an illustration for convenience: In Figure \ref{fig:leadtime}, the customers first choose the lead time option. The price shown on this page is the cheapest price for that lead time option (among all time-window options). In Figure \ref{fig:shipwindow}, customers can choose the exact time window they want, which we call the second-level choice. 
For modeling, we can bucket the 24 hours of potential time windows into 12 buckets of 2 hours each, 8 buckets of 3 hours each, 6 buckets of 4 hours each, or 4 buckets of 6 hours each. This modeling choice depends on business needs and actual customer choices we see from real data.

More generally, we need to instantly quote prices to customers for $L$ different lead time options, and within each lead time option, $M$ different second-level options corresponding to time windows. In this section, we discuss a solution to extend the base model framework that provides all $LM$ prices at the time of price quoting quickly. Our novel contribution is that we do not generate $LM$ product-options and use the same ideas as before. Rather, we train another MNL model over $M$ options that aligns with the base model, can be implemented as an add-on, and can be adjusted as the offered set of time windows changes over time.

Let $\mathbf{x}$ denote the vector of features describing a current customer quote request. Let $c_{ij}$ be the cost of lead time option $i$ with option $j$. The goal is to choose a set of prices $p_{ij}$ to maximize a given objective, given the constraint that the customer sees the cheapest second level time-window option for each lead time option $i$ on the first page whose price has already been calculated using the base model framework and cannot be changed on the second page.  We elaborate below.

First, we obtain the base model conversion probability vector for the $L+1$ lead-time options (including no-choice) as well as the calendar date view lowest prices $p_i$, using the lowest cost $c_i=\min_j c_{ij}$ for each lead-time option. See previous sections for details. Let $Y_1$ denote the lead-time choice and $Y_2$ denote the time-window choice. For any given lead-time option $i$, we then create a new second-level MNL model for choosing time-window option $j: 1\leq j \leq M$ is defined as follows:
\begin{align*}
    \mathbb{P}(Y_2(\mathbf{p})=j|X=x, \textrm{clicked on } i)= \frac{e^{V_{ij}}}{1+ \sum_{k=1}^M e^{V_{ik}}}.
\end{align*}
In the equation above,  we have $V_{ij} = \alpha_j - \beta p_{ij} + \mathbf{\gamma}^T \mathbf{x}_i  + \delta j + \epsilon_{ij},$
$\alpha_j$ is constant, $\beta$ is the common price sensitivity for time-window options, 
$\gamma$ is the coefficient for the quote features,  $\delta$ is the coefficient for the second-level option, and $\epsilon_{ij}$ are i.i.d. Gumbel noise (similar to the standard assumption for discrete choice models resulting in an MNL choice model).

\textbf{Using existing data to fit MNL model} We have detailed data on costs/prices for each time-window option. For the converted quotes, we know which specific lead-time and time-window option combination was chosen. In our dataset, each row corresponds to one quote (and vice-versa, i.e., each quote is represented in one row only) with details about the prices of all time-window options and the final choice of the customer (no purchase or a specific time-window). For converted quotes, we can simply use the chosen lead-time as the single set of prices for time-window options and other features. However, for unconverted quotes, we do not know which lead-times the customer clicked on (if any). 
Here, we take advantage of the lead-time specific conversion probability vector that we can obtain from the base model, which is already known at this point, to impute and subsample this missing feature for the no-conversion rows. 

\textbf{Price optimization} Once we have fitted the above model, we can calculate the conversion probability vector over all the different time-window options (including the no-purchase option) for any given lead-time and prices. Since we already know the predicted costs for each time-window option as well, we can easily set up the price optimization problem. Note that a one-parameter cost-adjusted markup pricing policy is optimal for the second-level MNL model \cite{talluri2004revenue}. Hence, this price optimization is a simple one-dimensional search for the optimal markup with the constraint that the prices are consistent with the prices on the lead time selection page (the calendar view).

\textbf{A/B testing performance} We now describe the performance of our model and pricing solution with respect to the specific objective of overall profit (which is a special case of our generalized objective), which can be computed using
\begin{equation*}
    \begin{aligned}
     \sum_{i=1}^L \frac{\mathbb{P}(Y_1(\mathbf{p})=i|X=x)}{1-\mathbb{P}(Y_1(\mathbf{p})=0|X=x)} &  \sum_{j=0}^M (p_{ij}-c_{ij}) \big\{ \\ 
    & \mathbb{P}(Y_2(\mathbf{p})=j|X=x,\textrm{clicked on } i) \big\},
\end{aligned}
\end{equation*}
where the term $\mathbb{P}(Y_1(\mathbf{p})=i|X=x)$ comes from the base model and $\mathbb{P}(Y_2(\mathbf{p})=j|X=x,\textrm{clicked on } i)= \frac{e^{V_{ij}}}{1+\sum_{k=1}^M e^{V_{ik}}}$ comes from the MNL model. The first term readjusts the purchase probabilities of the base model to account for the fact the no-purchase option should only be factored in once, and that occurs in the second-level model. In other words, we assume every customer goes to the second level, even if they chose not to purchase anything. Without this assumption, it is difficult to calibrate the no-purchase probability precisely.

Using another A/B testing pilot, our methodology improved performance significantly on the order of 4-10\% with respect to quotes utilizing the second level time window options. This second-level model was fully deployed and achieved a wide-ranging success in the organization.

\balance

\section{Conclusion}\label{sec:conc}
This paper discusses and solves the primary problem of pricing live service quote requests for customers in a business setting where they are presented with a multitude of scheduled service options, often in an hierarchical manner. Our proposed approach leverages a combination of parametric and non-parametric tools. Specifically, we leverage Market Segmentation Trees alongside the reference-price effects MNL choice model. Our pricing algorithm leverages known optimality structure to quickly approximate a high-dimensional optimization problem. We found that our new method offered significant improvements in real A/B tests, while simultaneously improving prediction accuracy, price stability, and pricing quote latency. Moreover, our new approach was amenable to addressing the pricing of second-level time window options, without exploding the number of model parameters to be estimated. Future research directions will be to consider how to more directly incorporate reinforcement learning for tackling price exploration and network optimization tools into the price optimization pipeline.

\bibliographystyle{plainnat}
\bibliography{bibly}

@article{aouad2023market,
  title={Market segmentation trees},
  author={Aouad, Ali and Elmachtoub, Adam N and Ferreira, Kris J and McNellis, Ryan},
  journal={Manufacturing \& Service Operations Management},
  year={2023}
}

@book{gallego2019revenue,
  title={Revenue management and pricing analytics},
  author={Gallego, Guillermo and Topaloglu, Huseyin and others},
  volume={209},
  year={2019},
  publisher={Springer}
}

@article{aouad2022representing,
  title={Representing random utility choice models with neural networks},
  author={Aouad, Ali and D{\'e}sir, Antoine},
  journal={arXiv preprint arXiv:2207.12877},
  year={2022}
}

@article{wang2024neural,
  title={Neural-network mixed logit choice model: Statistical and optimality guarantees},
  author={Wang, Zhi and Gao, Rui and Li, Shuang},
  journal={Available at SSRN 5118033},
  year={2024}
}

@inproceedings{ye2018customized,
  title={Customized regression model for airbnb dynamic pricing},
  author={Ye, Peng and Qian, Julian and Chen, Jieying and Wu, Chen-hung and Zhou, Yitong and De Mars, Spencer and Yang, Frank and Zhang, Li},
  booktitle={Proceedings of the 24th ACM SIGKDD international conference on knowledge discovery \& data mining},
  pages={932--940},
  year={2018}
}

@article{cai2022deep,
  title={Deep learning for choice modeling},
  author={Cai, Zhongze and Wang, Hanzhao and Talluri, Kalyan and Li, Xiaocheng},
  journal={arXiv preprint arXiv:2208.09325},
  year={2022}
}

@article{chen2022decision,
  title={Decision forest: A nonparametric approach to modeling irrational choice},
  author={Chen, Yi-Chun and Mi{\v{s}}i{\'c}, Velibor V},
  journal={Management Science},
  volume={68},
  number={10},
  pages={7090--7111},
  year={2022},
  publisher={INFORMS}
}

@article{farias2013nonparametric,
  title={A nonparametric approach to modeling choice with limited data},
  author={Farias, Vivek F and Jagabathula, Srikanth and Shah, Devavrat},
  journal={Management science},
  volume={59},
  number={2},
  pages={305--322},
  year={2013},
  publisher={INFORMS}
}

@article{fisher2014demand,
  title={A demand estimation procedure for retail assortment optimization with results from implementations},
  author={Fisher, Marshall and Vaidyanathan, Ramnath},
  journal={Management Science},
  volume={60},
  number={10},
  pages={2401--2415},
  year={2014},
  publisher={INFORMS}
}

@article{wang2012capacitated,
  title={Capacitated assortment and price optimization under the multinomial logit model},
  author={Wang, Ruxian},
  journal={Operations Research Letters},
  volume={40},
  number={6},
  pages={492--497},
  year={2012},
  publisher={Elsevier}
}

@article{hopp2005product,
  title={Product line selection and pricing with modularity in design},
  author={Hopp, Wallace J and Xu, Xiaowei},
  journal={Manufacturing \& Service Operations Management},
  volume={7},
  number={3},
  pages={172--187},
  year={2005},
  publisher={INFORMS}
}

@book{train2009discrete,
  title={Discrete choice methods with simulation},
  author={Train, Kenneth E},
  year={2009},
  publisher={Cambridge university press}
}

@article{besbes2020pricing,
  title={Pricing analytics for rotable spare parts},
  author={Besbes, Omar and Elmachtoub, Adam N and Sun, Yunjie},
  journal={INFORMS Journal on Applied Analytics},
  volume={50},
  number={5},
  pages={313--324},
  year={2020},
  publisher={INFORMS}
}

@inproceedings{elmachtoub2017practical,
  title={A Practical Method for Solving Contextual Bandit Problems Using Decision Trees},
  author={Elmachtoub, Adam N and McNellis, Ryan and Oh, Sechan and Petrik, Marek},
  year={2017},
  booktitle={{UAI} '17: Proceedings of the 33rd Conference on Uncertainty in Artificial Intelligence}
}

@article{wang2018prospect,
  title={When prospect theory meets consumer choice models: Assortment and pricing management with reference prices},
  author={Wang, Ruxian},
  journal={Manufacturing \& Service Operations Management},
  volume={20},
  number={3},
  pages={583--600},
  year={2018},
  publisher={INFORMS}
}

@article{talluri2004revenue,
  title={Revenue management under a general discrete choice model of consumer behavior},
  author={Talluri, Kalyan and Van Ryzin, Garrett},
  journal={Management Science},
  volume={50},
  number={1},
  pages={15--33},
  year={2004},
  publisher={INFORMS}
}

@article{garrow2007much,
  title={How much airline customers are willing to pay: An analysis of price sensitivity in online distribution channels},
  author={Garrow, Laurie A and Jones, Stephen P and Parker, Roger A},
  journal={Journal of Revenue and pricing Management},
  volume={5},
  pages={271--290},
  year={2007},
  publisher={Springer}
}

\end{document}